\documentclass[11pt,twoside]{article}
\usepackage{asp2010}

\resetcounters

\begin{document}

\title{Exploring the Morphology of the Expanding Remnants of Classical and Recurrent Novae}
\author{V. A. R. M. Ribeiro$^1$, M. F. Bode$^1$, M. J. Darnley$^1$, U. Munari$^{2,3}$, D. J. Harman$^1$
\affil{$^1$Astrophysics Research Institute, Liverpool John Moores University, Twelve Quays House, Egerton Wharf, Birkenhead, CH41 1LD, UK}
\affil{$^2$INAF Astronomical Observatory of Padova, via dell'Osservatorio, 36012 Asiago (VI), Italy}
\affil{$^3$ANS Collaboration, c/o Astronomical Observatory, 36012 Asiago (VI), Italy}}

\begin{abstract}
We report studies of several novae which are known or suspected to be recurrent. We discuss our morpho-kinematical modelling of the evolution of the optical spectra taken early after outburst for two recent novae. In the case of the known RN RS Oph, this is also coupled with {\it HST} imaging. Results support the hypothesis that remnant shaping occurs very early in a nova outburst and we also derive the structures (including inclination) and velocity field of the remnants. Overall, these results emphasise the need for coordinated imaging and spectrometry, although not always possible, if we are to truly understand remnant shaping in these systems, together with the wider implications for studies of shaping mechanisms in Planetary Nebulae.\\
\noindent{\bf Keywords.}\hspace{10pt}Planetary Nebulae -- Novae
\end{abstract}

\section{Introduction}
The classical nova (CN) outburst is due to a thermonuclear runaway on the surface of a white dwarf \citep[WD; see e.g.,][these proceedings]{BE08,E10} occurring within the matter accreted from a close companion star. A related class of objects are the recurrent novae (RNe), which have high mass WDs, probably close to the Chandrasekhar limit, to account for the short recurrence time-scales, and high accretion rates \citep[e.g.,][]{SST85,YPS05}.

Classical nova remnants show a range of morphologies whose basic origin has been modelled via a combination of a common envelope phase and binary motion \citep[e.g.,][and references therein]{POB98}. In the case of RN RS Oph-type remnants, shaping is suggested to be due to interaction of the ejected material with the pre-existing red-giant wind.

This paper will focus on our work with morpho-kinematical modelling, using Shape \citep[see poster 32;][]{SL06,SKW10} of three novae to determine their 3D morphology (including inclination) and velocity field with a combination of {\it Hubble Space Telescope} ({\it HST}) imaging and ground-based spectrometry for RS Oph (Section \ref{rs_oph}) and in the case of V2491 Cyg and V2672 Oph using solely H$\alpha$ line profiles (Section \ref{others}). Finally in Section \ref{conclusions} we summarise our findings.

\section{The Expanding Nebular Remnant of the 2006 Outburst of RS Ophiuchi}\label{rs_oph} 
RS Oph is one of the most well studied RNe, with six previous recorded outbursts and two further ones suggested \citep[e.g.][]{R87,RI87,OM93,S04,EBO08}. From its latest outburst, beginning 2006 February 12.83 \citep{NHK06}, RS Oph was quickly followed with a multi-frequency campaign from the radio to X-ray \citep[e.g.,][]{SLM06,OBP06,BOO06,CNM07}.

\begin{figure}[ht]
\centering
\includegraphics[width=107mm]{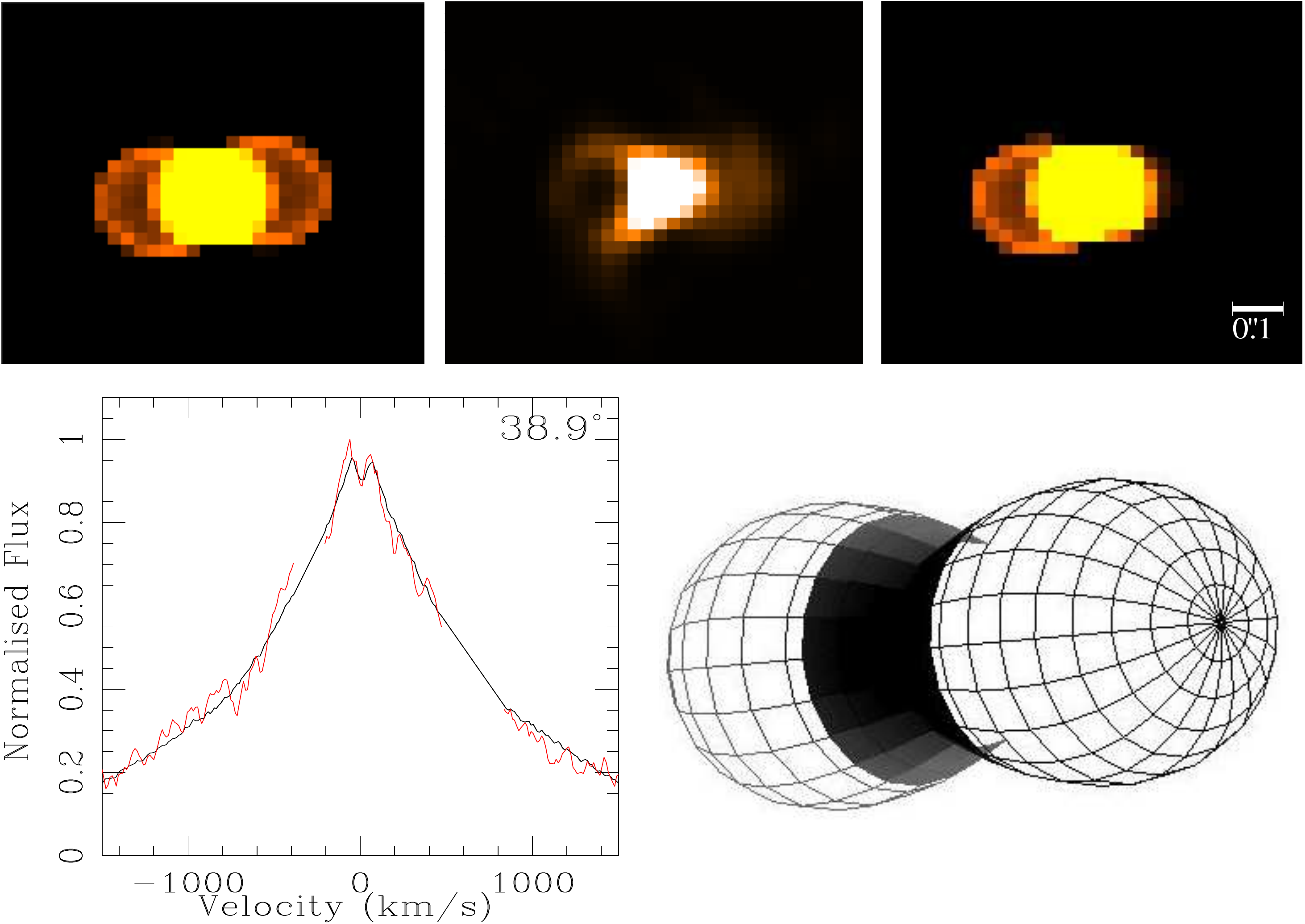}
\caption{Top: synthetic image without the {\it HST} F502N ACS/HRC filter profile applied (left), enlarged ACS/HRC image at $t$ = 155 days after outburst (middle) and synthetic image with the ACS/HRC F502N filter profile applied (right). Bottom: best fit synthetic spectrum (black) overlaid with the observed spectrum (red). To the right is the model structure for RS Oph (outer dumbbell and inner hour glass). Images reproduced from \citet{RBD09}.}
\label{fig:rs_oph}
\end{figure}

RS Oph was observed at around 155 days after outburst with {\it HST} imaging and ground-based spectrometry \citep{RBD09}. These observations were used to derive the physical parameters of the object. Figure \ref{fig:rs_oph} shows the results of this modelling, constraining a model with an outer dumbbell and high density inner hour glass structure. The higher density was required to replicate the lower expansion velocities observed in the spectrum. The inclination of the system was derived to be 39$^{+1}_{-10}$ degrees, comparable to estimates of the inclination of the central binary of \citet{DK94}. At a distance of 1.6$\pm$0.3 kpc (\citealt{B87}; see also \citealt{BMS08}) the implied maximum expansion velocity of the system is 5100$^{+1500}_{-100}$ km s$^{-1}$ (the range in velocity arise from the 1$\sigma$ errors on the inclination). The observed apparent asymmetry of the ejecta in the ACS/HRC image is suggested to be due to the effect of the finite width and offset from the [O III] line's rest wavelength of the F502N filter. The model was then evolved to 449 days after outburst (the second {\it HST} epoch). The results implied that the outer dumbbell structure expanded linearly while the inner hour glass structure shows some evidence for deceleration. However, since only lower quality WFPC2 imaging was available at that time, the results are open to over-interpretation.

\section{Early Spectrometric Evolution of V2491 Cygni and V2672 Ophiuchi}\label{others}
V2491 Cyg and V2672 Oph are two particularly interesting novae that have been suggested to be RNe candidates and that underwent their outbursts on 2008 April 10.8 \citep{NBJ08} and 2009 August 16.5 \citep{NYK09} respectively. V2491 Cyg is only the second nova with recorded pre-outburst X-ray observations \citep[after V2487 Oph - a RN; e.g.,][]{POE10}. V2672 Oph shows very similar characteristics to the RN U Sco. It displayed a photometric plateau phase, a He/N spectrum classification, extreme expansion velocities and triple peaked emission line profiles during the decline phase \citep{MRB10}.

\begin{figure}[ht]
\centering
\includegraphics[width=131mm]{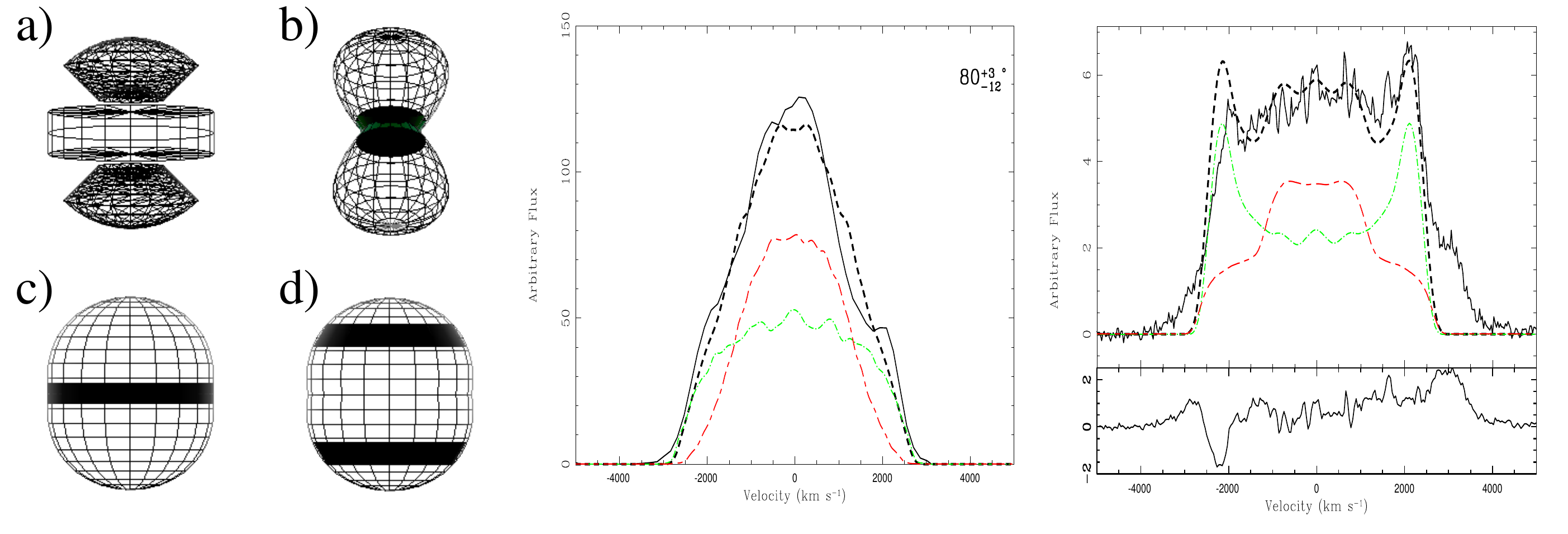}
\caption{Model fit to V2491 Cyg. Left: model structures used to replicate the emission lines; a) polar blobs with an equatorial ring; b) outer dumbbell with an inner hour glass; c) prolate shell with tropical rings; and d) prolate shell with an equatorial ring. Middle: best fit synthetic spectrum (dashed black) to the observed spectrum (solid black) on day 25 after outburst for the polar blobs (short-dash-long-dash red) and equatorial ring (dot-dashed green) structure. Right: observed spectrum at 108 days after outburst compared with the evolved model spectrum.}
\label{fig:v2491}
\end{figure}

Using the Meaburn Spectrograph on the Liverpool Telescope \citep{SSR04} we observed V2491 Cyg from 2 to 31 days after outburst and with the Asiago-AFOSC at 108 days after outburst. With the lack of resolved imaging to complement our spectra we first assumed some structures which have been previously observed in CNe (Figure \ref{fig:v2491}). We modelled the H$\alpha$ emission line at 25 days after outburst and then evolved our models to the later date (day 108). The effects of optical depth were investigated and we found that the line shape did not change considerably. We found that two models produced similar results at 25 days after outburst (polar blobs and equatorial ring and the dumbbell with an hour glass). However, when both models were evolved to day 108 we found that due to the contribution of [N II], which was not detected at the earlier date, the best-fit model was that for the polar blobs and an equatorial ring. We suggest that the inclination angle of the system is 80$^{+3}_{-12}$ degrees and a maximum expansion velocity of 3100$^{+200}_{-100}$ km s$^{-1}$  \citep[see][for further details]{RDB10}.

Similarly, we applied the techniques above to V2672 Oph and found that the best fit model was for polar blobs and an equatorial ring with an underlying prolate structure whose density appears to decline faster with time than that of the other components. The inclination of the system is suggested to be 0$\pm$6 degrees and a maximum expansion velocity of the polar blobs of 4800$^{+900}_{-800} $ km s$^{-1}$ \citep{MRB10}.

\section{Conclusions}\label{conclusions}
We have investigated the evolution of three novae to derive their 3D geometry (including inclination) and expansion velocities. We show that the best way to have a true grasp of these parameters is with combined multi-epoch imaging and spectrometry. Caution should be exercised when dealing solely with spectrometric data and any model should ideally have predictions. Useful lessons may be learnt from the study of remnant shaping in novae that may be applied to that in PNe \citep{E10}, particularly of course those with binary star nuclei.

\acknowledgements VARMR would like to thank financial support from the Royal Astronomical Society, STFC, the LOC and the Astrophysics Research Institute, LJMU.

\bibliographystyle{asp2010}

\end{document}